# AGF-Net: A Robust Multimodal Emotion Recognition Model for Asynchronous Audio-Visual Signals Based on Predictive Dynamic Fusion with Negative Learning

Chuan-Han Li[1]*, Yun-Chien Cheng[1]*,

[1] Department of Mechanical Engineering, College of Engineering, National Yang Ming Chiao Tung University, Hsin-Chu, Taiwan

[1] Department of Electrical Engineering, College of Engineering, National Taiwan University, Taipei, Taiwan

*Corresponding author: LCH20020726.en13@nycu.edu.tw, yccheng@ntu.edu.tw

## Abstract

This study aims to design an emotion recognition system that integrates automatic utterance segmentation with audio-visual multimodal fusion. The proposed system automatically generates candidate emotional segments from continuous dialogue data and further extracts the corresponding audio and visual features for emotion classification, thereby improving the applicability of emotion recognition models in real-world human-computer interaction scenarios. Previous studies on multimodal emotion recognition mostly use manually annotated utterance boundaries as model inputs. However, in real-world applications, systems usually cannot obtain the start and end times of utterances in advance and must automatically determine valid segments from continuous audio-visual signals. In addition, automatically segmented clips are easily affected by boundary shifts, overly short or long segments, speaker leakage, and unstable modality quality, which may result in incomplete or inconsistent audio and visual features and further reduce the stability of subsequent emotion recognition models. Based on these issues, this study establishes an automatic utterance segmentation process to divide continuous dialogue audio into emotional segments for downstream models and proposes the AGF-Net multimodal emotion recognition framework. This framework integrates three strategies: Audio-Guided Fusion, Predictive Dynamic Fusion, and Multimodal Negative Learning. Experimental results show that the proposed AGF-Net achieves 64.92% Accuracy and 65.47% F1-score under the clean setting, maintains 62.13% Accuracy and 63.08% F1-score under moderate noise, and reaches 58.19% Accuracy and 59.41% F1-score under high-intensity noise. Compared with existing multimodal emotion recognition methods, the proposed method demonstrates better overall performance under both clean and noisy conditions. In summary, this study provides an audio-visual emotion recognition pipeline that is closer to real-world human-computer interaction scenarios, effectively reduces the model's reliance on manually annotated utterance boundaries, and improves recognition robustness under automatically segmented clips and unstable modality quality.

## 1. Introduction

Emotion is a subjective experience elicited by external stimuli and plays a critical role in self-understanding and decision-making processes. In recent years, with the rapid growth of the internet-driven era, emotion recognition technologies have been increasingly adopted across diverse domains, including online learning, social media analytics, and human–robot interaction[1-3]. Therefore, developing accurate and reliable methods for emotion recognition is of substantial importance.

Human emotional expression is inherently complex and highly context-dependent, and is often manifested through multiple cues simultaneously, including prosody and speech characteristics, facial expressions, body gestures, textual content, and physiological signals[4-7]. Relying solely on a single modality for emotion recognition is therefore prone to error, as it is vulnerable to inter-individual variability, environmental noise, and missing or degraded data, and may yield misclassifications due to incomplete information.

However, most existing multimodal emotion recognition approaches rely on manually segmented utterances with predefined temporal boundaries. In real-world applications, systems typically receive continuous audio-visual streams, in which the beginning and end of each utterance are unknown. Therefore, an applicable emotion recognition system must not only integrate information from multiple modalities but also automatically identify meaningful segments from continuous input signals.[2-7]

To address these challenges, this study proposes an audio-visual emotion recognition system that integrates automatic utterance segmentation with multimodal feature fusion. The proposed system first detects candidate utterance segments from continuous audio-visual streams. The corresponding audio and visual information is then extracted from each segment and processed by modality-specific feature extractors. Finally, the extracted features are fused to perform emotion classification. By reducing the reliance on manually annotated utterance boundaries and exploiting complementary information across modalities, the proposed system aims to maintain robust recognition performance even when the automatically generated segments are imperfect or when one modality is insufficient or affected by interference. This design enhances the applicability and reliability of emotion recognition in real-world human–robot interaction scenarios.

In recent years, a growing body of research has leveraged multimodal information for emotion recognition. For instance, Wang et al[8]. proposed a CNN-based emotion recognition framework that integrates visual and physiological signals to improve emotion classification performance; Salas et al[9]. adopted an LSTM-centered approach to model the temporal dynamics of visual and audio sequences for emotion recognition; and Khan et al[10]. further employed a Transformer architecture to fuse audio and textual modalities for emotion analysis and classification. These studies collectively indicate that multimodal fusion can strengthen emotion inference by exploiting complementary cues across modalities.

However, applying multimodal emotion recognition to continuous real-world audio-visual streams presents four key challenges:

(1) Dependence on predefined utterance boundarie**s:** Most existing approaches assume that the beginning and end of each utterance are known in advance and use manually segmented samples as model inputs. In real-world applications, however, systems typically receive continuous audio-visual signals without predefined utterance boundaries. Therefore, meaningful segments must be automatically detected before feature extraction and emotion classification. Inaccurate segmentation may produce incomplete, excessively short, or overly long segments, thereby degrading the quality of the extracted features and subsequent recognition performance.

(2) Modality imbalance: During multimodal training, the model may overly rely on a single, easier-to-learn or more informative modality, causing other modalities to be under-utilized and systematically down-weighted, thereby limiting the benefits of cross-modal fusion; differences in signal quality, missing modalities, or noise can further exacerbate this phenomenon.

(3) Temporal misalignment: Due to mismatches in sampling rates, timestamps, or response latencies across modalities (e.g., temporal offsets between speech and facial expressions, or misalignment between text and speech), insufficient alignment can weaken cross-modal complementarity and reduce robustness.

(4) Lack of effective fusion strategies: Simple feature concatenation or fixed-weight fusion often fails to capture inter-modality interactions. It cannot dynamically adjust each modality's contribution according to its reliability, which ultimately constrains overall recognition performance.

In summary, this study aims to develop a multimodal emotion recognition model that can effectively integrate information across modalities. To address the reliance on predefined utterance boundaries in continuous audio-visual streams, existing automatic segmentation strategies can be broadly categorized into three types. First, time-based methods [11,12] employ voice activity detection and predefined silence thresholds to determine utterance boundaries, offering low computational cost and real-time applicability but remaining sensitive to intra-sentence pauses. Second, model-based methods [13,14] learn boundary decisions from acoustic or contextual patterns, providing greater flexibility but requiring annotated training data and higher computational complexity. Third, semantic-assisted methods [15,16] utilize linguistic and semantic cues, such as punctuation, contextual continuity, and sentence completeness, to determine whether an utterance has been completed, although their performance depends on the reliability of the extracted semantic information. Considering the trade-offs among latency, segmentation accuracy, and implementation cost, this study combines voice activity detection(VAD) with semantic assistance to generate candidate segments and refine their boundaries before multimodal emotion recognition.

To address the common issue of temporal asynchrony and cross-modal misalignment in multimodal data, existing strategies can be broadly categorized into three types. First, alignment-oriented[17-19] methods perform temporal warping or re-parameterization of modality-specific timelines to reduce cross-modal alignment offsets and improve synchronization. Second, scale-aware temporal modeling[20-21] captures feature dynamics at multiple temporal scales to accommodate potential response delays and peak discrepancies across modalities during emotional expression. Third, anchor-based[22-23] method**s** designate a modality with higher signal quality or stronger discriminative power as an anchor to guide temporal alignment and feature fusion for the remaining modalities. Given the characteristics of our dataset and application requirements, this work adopts an anchor-based strategy as the core approach to enhance the robustness of cross-modal alignment and improve overall recognition performance.

Prior studies have adopted an "anchor based" paradigm to guide cross-modal alignment and fusion. For example, Shi et al. [22] used audio as the anchor modality to further steer the integration of visual and textual information, thereby improving the discriminative power of multimodal representations; Li et al. [23] performed cross-modal knowledge transfer via knowledge distillation, transferring discriminative cues from a strong modality to weaker modalities to mitigate modality dominance and enhance overall performance. In addition, Shi et al. [24] proposed a text-anchored cross-slice scoring module that dynamically assigns attention weights to each 2D slice conditioned on task-specific instructions and subsequently combines the resulting representations with 3D features to strengthen representation learning. Building on these anchor-based approaches, this study will draw on and extend the paradigm to improve the performance of multimodal emotion recognition models.

To enhance the complementarity of multimodal features, this study integrates Hybrid Fusion into a deep learning model to strengthen cross-modal information integration. Prior studies, such as Wang et al. [25] and Wolters et al. [26], have shown that hybrid fusion can improve the overall performance of multimodal approaches. Inspired by these findings, we adopt a hybrid fusion strategy to enhance model performance.

In addition, to mitigate modality imbalance in multimodal learning, Xu et al[27]. identified four areas for improvement: Data, Feedforward, Objective, and Optimization. At the data level, selective resampling can increase learning opportunities for weaker modalities, but the processing pipeline is relatively complex. At the feed-forward level, modifying feature processing or fusion modules can encourage better utilization of weaker modalities. At the objective level, redesigning the loss function or learning targets can suppress modality dominance. At the optimization level, gradient regulation can reserve more optimization capacity for weaker modalities, but it usually incurs higher computational cost. Considering feasibility and resource constraints, this study focuses on improvements at the Feed-forward and Objective levels to reduce the impact of modality imbalance on model performance.

In summary, this study proposes the AGF-Net architecture to enhance multimodal emotion recognition under continuous audio-visual conditions. The proposed framework integrates automatic utterance segmentation with multimodal feature fusion to reduce the reliance on predefined utterance boundaries. It further alleviates three major limitations of conventional multimodal models, including insufficient cross-modal feature complementarity, modality imbalance, and temporal misalignment. Compared with existing methods, the proposed approach achieves improved recognition performance and demonstrates the potential for broader real-world applications. The main contributions of this study are summarized as follows:

(1) Developing an automatic utterance segmentation strategy to extract meaningful segments from continuous audio-visual streams without relying on predefined utterance boundaries.

(2) Strengthening cross-modal feature complementarity and fusion.

(3) Mitigating modality imbalance and modality dominance.

(4) Improving recognition robustness under imperfect segmentation, noisy, missing, or temporally misaligned conditions.

## 2. Related Work

### 2.1. Wav2vec 2.0

Wav2vec2.0[28] is a self-supervised speech representation learning model that learns transferable high-level features directly from raw waveforms and can be applied to ASR and various speech understanding tasks. Its architecture typically consists of three components: a 1D CNN feature encoder that converts the waveform into latent representations, a Transformer-based context network that models long-range temporal dependencies to produce contextualized features, and a quantization module that discretizes continuous representations to provide more stable learning targets.

During pre-training, wav2vec 2.0 masks a subset of time steps and optimizes a contrastive objective, requiring the model to identify the correct target representation from negative samples based on the context representation. This design enables effective learning from large-scale unlabeled speech data and improves downstream performance after fine-tuning. Overall, wav2vec 2.0 has become a widely used benchmark in speech representation learning due to its reduced reliance on labeled data.

### 2.2. ResNet18

ResNet18[29] is a widely used deep convolutional neural network in computer vision. Its key strength lies in the residual connection design, which stabilizes optimization, improves convergence, and alleviates the performance degradation that can occur as network depth increases. As a result, ResNet18 can learn highly discriminative and generalizable visual representations without relying on overly complex architectures.

Owing to its strong feature extraction capability and mature training/deployment ecosystem, ResNet18 is commonly adopted as a backbone in a wide range of vision tasks, including image classification and feature embedding, and as the feature encoder for object detection and image segmentation models. Overall, it offers a well-balanced trade-off between accuracy, robustness, and computational cost, making it a standard baseline and a frequent point of comparison in both research and practical applications.

### 2.3. Silero VAD

Silero VAD is a lightweight neural voice activity detection model designed to identify speech and non-speech regions from continuous audio signals. Unlike traditional VAD methods that mainly rely on handcrafted acoustic features or energy-based thresholds, Silero VAD uses a deep learning-based architecture to estimate the probability of speech activity for each short audio frame. Due to its low computational cost and stable detection performance, it is suitable for real-time speech processing tasks such as speech segmentation, endpoint detection, and automatic utterance extraction.

During inference, the input waveform is divided into short streaming chunks, and Silero VAD outputs a speech probability for each chunk. These probabilities can then be combined with smoothing strategies and threshold-based decision rules to determine whether the current audio state corresponds to speech or silence. In this study, Silero VAD is used as the foundation of the automatic utterance segmentation process. By detecting speech onset and endpoint from continuous dialogue audio, it enables the system to generate candidate emotional segments without relying on manually annotated utterance boundaries. Overall, Silero VAD provides an efficient and practical solution for real-time audio segmentation and serves as an important preprocessing module for the subsequent audio-visual emotion recognition model.

### 2.4. WhisperX

WhisperX [xx] is an automatic speech recognition and alignment framework extended from the Whisper model. It is designed to generate speech transcriptions while providing more accurate temporal information, especially at the word level. Compared with conventional ASR systems that often provide only segment-level timestamps, WhisperX combines speech recognition with voice activity detection and forced alignment, allowing the transcribed words to be mapped more precisely to their corresponding positions in the original audio.

The overall process of WhisperX typically consists of three main stages. First, the input audio is segmented into speech regions using voice activity detection, which reduces the difficulty of processing long-form audio and improves transcription efficiency. Second, the segmented speech regions are transcribed by the Whisper ASR model to obtain the corresponding text. Finally, an additional alignment model is applied to align the recognized text with the original waveform, producing more accurate word-level timestamps. This design enables WhisperX to provide both textual content and fine-grained temporal information, making it useful for tasks that require speech transcription, temporal alignment, and utterance boundary analysis.

## 3. Material and Methods

### 3.1. Dataset

This study uses the publicly available IEMOCAP (Interactive Emotional Dyadic Motion Capture) dataset [30], a widely used multimodal emotion recognition dataset collected from dyadic human interactions. The dataset contains approximately 12 hours of audio–visual recordings from 10 actors across five sessions, including both scripted and improvised conversations. Each utterance is annotated with categorical emotion labels, making the dataset suitable for evaluating emotion recognition models under conversational scenarios. In this work, we focus on four commonly used emotion categories, including angry, happy, neutral, and sad. The corresponding audio and visual data are used as model inputs, while the categorical emotion labels serve as the learning targets. The dataset is further divided into training, validation, and test sets for model training and performance evaluation.

### 3.2. Experimental Procedure

As shown in Figure 1, this study first collects the data and divides it into training, validation, and test sets. Next, automatic utterance segmentation is performed on continuous dialogue data to generate candidate segments for subsequent emotion recognition, and the corresponding audio and visual features are further extracted. In terms of model design, this study proposes an audio-visual multimodal emotion recognition model suitable for automatically segmented clips. Finally, to validate the effectiveness of the proposed method, a series of experiments and ablation studies are conducted, including:

(1) Evaluating the performance of the automatic utterance segmentation strategy.

(2) Comparing the recognition performance of different fusion methods.

(3) Analyzing the model performance after incorporating the Audio-Guided Fusion (AGF) module.

(4) Investigating the effect of different warm-up epoch settings in Multimodal Negative Learning (MNL) on model performance.

(5) Analyzing the model performance and noise robustness after incorporating the Predictive Dynamic Fusion (PDF) module.

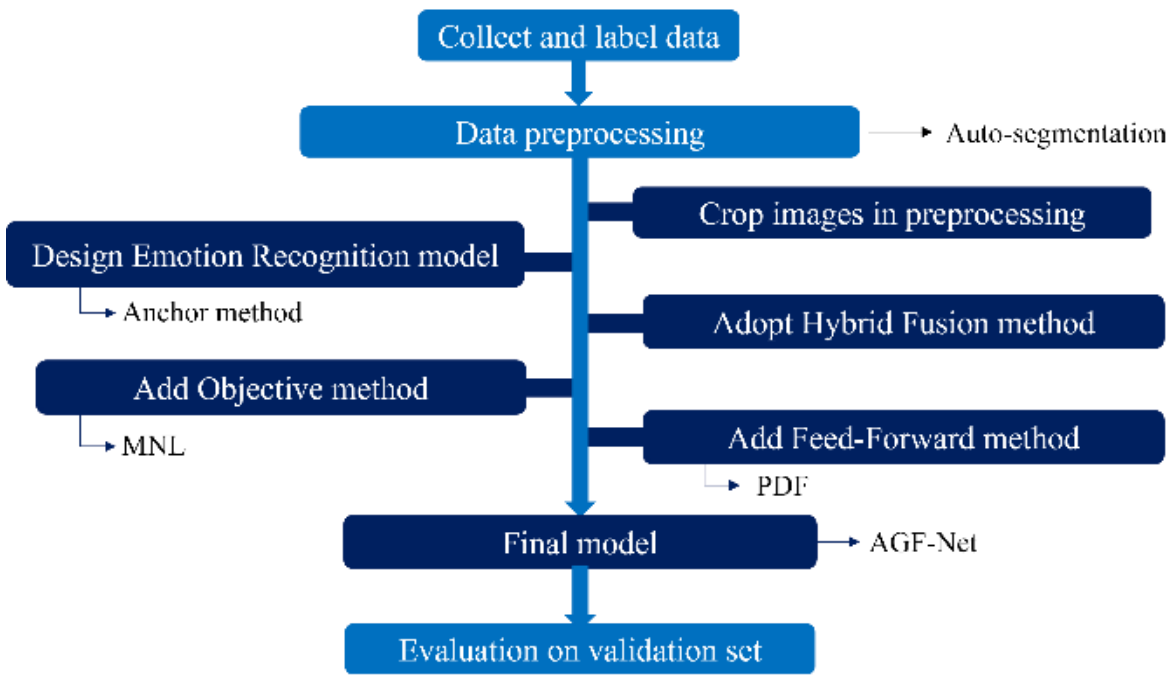

Figure 1. Experiment process

### 3.3. Data Preprocessing

The preprocessing pipelines for the visual and audio modalities are described in the following sections.

*3.3.1 Image Preprocessing*

For visual-modality preprocessing, this study employs YuNet [31] for face detection and cropping. The model jointly predicts a face bounding box and five facial landmarks (both eyes, the nose tip, and the mouth corners). We use these landmarks to perform face alignment, thereby reducing the interference of head-pose variations on emotion inference. After alignment and cropping, each face image is resized to 224×224 as the input to the subsequent ResNet18 feature extractor.

*3.3.2 Audio Preprocessing*

For audio-modality preprocessing, we first extract the speech signal from each video clip and resample it to 16 kHz to standardize the sampling rate and match the input requirement of wav2vec 2.0. Next, we apply z-score normalization to the audio waveform to reduce amplitude-scale variations caused by different recording conditions and inter-subject differences.

In addition, to prevent data leakage caused by the same actor appearing across the training, validation, and test sets—which could lead to an overestimation of model performance—we adopt an actor-ID-based subject-independent split strategy. Specifically, all clips from the same actor are assigned exclusively to a single split, enabling a more realistic evaluation of generalization to unseen subjects.

### 3.4. Automatic Utterance Segmentation

To address the issue that utterance boundaries are unavailable in real-world human-computer interaction scenarios, this study introduces an automatic utterance segmentation strategy to generate candidate emotional segments from continuous dialogue audio. Instead of relying on manually annotated utterance boundaries, the proposed process detects speech activity and determines endpoints from the audio stream, allowing the subsequent audio-visual emotion recognition model to operate on automatically segmented clips.

The core idea is to formulate utterance segmentation as a speech-state tracking problem. Given a continuous audio signal, the input is divided into short streaming chunks and processed by Silero VAD to obtain the speech probability $p_t$ at each time step. To reduce short-term prediction fluctuations, an exponentially weighted moving average is applied:

$$\tilde{p_t} = \alpha p_t + (1-\alpha)\tilde{p}_{t-1}$$

where $p_t$ is the original VAD speech probability, $\tilde{p_t}$ is the smoothed probability, and $\alpha$ is the smoothing coefficient. Based on the smoothed probability, a hysteresis threshold mechanism is used to determine whether the current state is speech or silence.

When a possible silence region is detected, an endpoint decision strategy is applied to determine whether the current segment should be emitted. In addition, adaptive endpoint adjustment is used to balance segmentation accuracy and response latency under different segment lengths. After the endpoint is determined, pre-roll and post-roll are added to preserve boundary information, and overly short segments are removed. Through this design, the proposed strategy can generate usable candidate clips for downstream audio-visual emotion recognition while reducing the dependence on ground-truth utterance boundaries.

### 3.5. Audio-Guided Lag-aware Network

To effectively address the asynchrony commonly observed between audio events and visual frames in multimodal sequences.The core idea is to formulate cross-modal semantic alignment as a probability density estimation problem constrained by a Gaussian prior: rather than relying solely on content-based matching, the model explicitly predicts the temporal center and duration of the corresponding visual signal for each audio event. Specifically, given an audio event sequence $E_a \epsilon \mathbb{R}^{K\times D}$ a visual feature sequence $F_v \epsilon \mathbb{R}^{V\times D}$, a lag predictor maps each audio event to$(\mu_i, \sigma_i)$, where $\mu_i$ denotes the location of the most relevant visual frame and $\sigma_i$ quantifies the alignment uncertainty/temporal span. We further introduce Gaussian-biased attention, which dynamically generates a Gaussian mask $G \in \mathbb{R}^{K\times V}$ and injects it as a bias term into scaled dot-product attention, so that attention weights are jointly constrained by content similarity and temporal priors. With this design, the model can, while maintaining interpretability, focus on key visual representations temporally consistent with the audio events and effectively suppress irrelevant background signals.

### 3.6. PDF

Cao et al [32] proposed PDF to handle open-world scenarios where modality quality varies across samples and over time, providing theoretically grounded and dynamically adjustable decision-level fusion weights. Starting from a generalization error upper bound, PDF derives the relationship between fusion weights and modality-specific losses, indicating that the weights should satisfy a collaborative condition: they should be negatively correlated with a modality's own loss while being positively correlated with the losses of other modalities. Accordingly, rather than regressing the typically unstable loss values, PDF uses the ground-truth class probability as a learnable confidence surrogate. Specifically, it first estimates a mono-confidence for each modality, then incorporates the states of other modalities to form a holo-confidence, and linearly combines them to obtain a collaborative belief (Co-Belief) that estimates the modality's contribution during fusion. Furthermore, to account for inevitable modality uncertainty in real deployments, PDF introduces Relative Calibration (RC), which quantifies uncertainty via the uniformity of the output probability distribution and applies asymmetric down-weighting to modalities with higher uncertainty. This yields more robust dynamic fusion weights, improving fusion reliability and stability under noise and modality-quality fluctuations.

### 3.7. MNL

Gong et al[33] proposed MNL under a decision-level fusion framework to alleviate fusion instability caused by modality imbalance and noise/quality fluctuations. Compared with prior approaches that rely on positive alignment to force weaker modalities to mimic stronger ones—potentially suppressing complementary information and propagating errors when the dominant modality fails—MNL adopts a negative learning strategy: a more reliable modality guides weaker modalities to suppress non-target classes, thereby stabilizing their decision space while preserving modality-specific cues. MNL dynamically identifies strong and weak modalities at the sample level and augments the standard cross-entropy objective with an uncertainty-suppression loss on non-target classes. Training typically follows a two-stage procedure, where the model is first warmed up with supervised objectives to ensure convergence of unimodal branches and the fusion module, and MNL is then activated to further improve fusion performance via negative learning. Since MNL introduces no additional computational overhead during inference, it is particularly suitable for the setting of this study.

### 3.8. Model Overview

The overall architecture of the proposed AGF-Net is illustrated in Figure 2. Given an automatically segmented audio-visual clip, the model first extracts audio and visual features, and then integrates the two modalities through the AGF module to strengthen the relationship between audio and visual emotional cues. At the decision stage, PDF dynamically assigns fusion weights according to the reliability of each branch, allowing more reliable branches to contribute more to the final prediction. In addition, MNL is introduced during training to stabilize the decision space of weaker branches by suppressing non-target classes, thereby alleviating modality imbalance. Overall, AGF-Net combines AGF, PDF, and MNL to improve the stability and robustness of audio-visual emotion recognition under automatically segmented inputs.

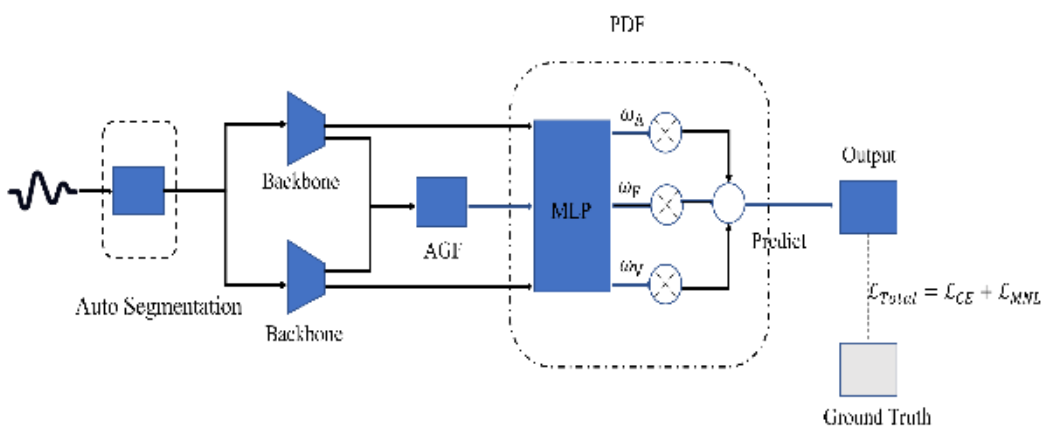

Figure 2 Overview of model

### 3.9. Equipment

The experiment uses an ASUS Z790-A GAMING WIFI 6E server equipped with an Intel i9-13900K processor and a single MSI RTX 4090 GAMING X TRIO 24G graphics card.

## 4. Result

### 4.1. Evaluation Methods

In this study, the performance of all deep learning architectures is evaluated using Accuracy, Precision, Recall, and F1-score. The metrics are computed as follows:

$$Accuracy = \frac{TP+TN}{TP+TN+FP+FN} \quad (1)$$

$$Precision = \frac{TP}{TP+FP} \quad (2)$$

$$Recall = \frac{TP}{TP+FN} \quad (3)$$

$$F1-scoree = 2 \times \frac{Precision \times Recall}{Precision + Recall} \quad (4)$$

In this evaluation framework, TP denotes the number of samples correctly predicted as the target class; TN denotes the number of samples correctly predicted as not belonging to the target class; FP denotes the number of samples incorrectly predicted as the target class while actually belonging to other classes; and FN denotes the number of samples that belong to the target class but are incorrectly predicted as other classes. Based on these definitions, Precision and Recall measure the correctness of positive predictions and the coverage of positive samples, respectively. The F1-score, as the harmonic mean of Precision and Recall, is further adopted to provide a comprehensive assessment of overall performance.

In addition to the classification metrics, Intersection over Union (IoU) is used to evaluate the quality of automatic utterance segmentation. IoU measures the temporal overlap between a predicted segment and its corresponding ground-truth utterance segment, and is defined as:

$$IoU = \frac{S_{pred} \cap S_{gt}}{S_{pred} \cup S_{gt}} \quad (5)$$

where $S_{pred}$ denotes the predicted segment generated by the automatic segmentation strategy, and $S_{gt}$ denotes the ground-truth utterance segment. A higher IoU value indicates that the predicted segment is temporally closer to the ground-truth boundary, while a lower IoU value suggests boundary shift, over-segmentation, or under-segmentation.

### 4.2. Implement Detail

The specific training configuration for the proposed model is as follows. The total number of training epochs was set to 100, and an early stopping mechanism was adopted to mitigate the risk of overfitting. AdamW was used as the optimizer to improve training stability and convergence. A layer-wise learning rate schedule was applied, where the learning rate was set to $1 \times 10^{-4}$ for the prediction head and $1 \times 10^{-5}$ for both the visual and audio branches. The batch size was set to 8.

For input settings, visual frames were detected, cropped, and resized to a fixed resolution of $224 \times 224$. For each automatically segmented clip, visual frames were extracted from the corresponding time interval to construct the visual input. Audio signals were extracted from the same segment boundaries and resampled to 16 kHz for subsequent feature extraction. Instead of truncating or padding all audio clips to a fixed duration, this study preserves the original duration of each automatically segmented segment, enabling the inputs to better reflect the temporal characteristics of real dialogue utterances.

Regarding the loss design, a composite objective function was employed, consisting of cross-entropy classification losses for the audio branch, the visual branch, and the fused branch, as well as a mean squared error (MSE) loss for reliability prediction. The loss weight for reliability prediction was set to 1.0, the auxiliary classification loss weight was set to 0.2, and the MNL loss weight was set to 1.0.

To further evaluate the robustness of the proposed model under unstable modality quality, noise injection was applied during the experiments. For each sample, noise was randomly added to either the audio modality or the visual modality, rather than perturbing both modalities simultaneously. This setting simulates a practical scenario in which only one modality is degraded while the other remains relatively reliable.

For the audio modality, additive white Gaussian noise (AWGN) was applied according to the RMS energy of the original signal:

$$\overline{x_a} = x_a + \frac{\lambda}{100} \cdot RMS(x_a) \cdot \epsilon \qquad \epsilon \sim N(0,1) \quad (6)$$

where $x_a$ denotes the original audio signal, $\overline{x_a}$ denotes the noisy audio signal.

For the visual modality, Gaussian noise was added to the image input as follows:

$$\overline{x_v} = x_v + \frac{\lambda}{255} \cdot \eta \qquad \eta \sim N(0,1) \quad (7)$$

where $x_v$ denotes the original visual input, $\overline{x_v}$ denotes the noisy visual input. In this study, $\lambda$ controls the noise intensity.

### 4.3. Automatic Segmentation

This section evaluates the performance of the proposed automatic utterance segmentation strategy. As shown in Table 1, the confidence endpoint strategy achieves an F1-score of 0.4372 under the IoU (>0.5) criterion, with an average endpoint latency of 546.8 ms. When the adaptive endpoint strategy is applied, the F1-score improves to 0.4532, indicating that dynamically adjusting the endpoint decision according to segment length and VAD confidence can produce better segmentation quality.

Furthermore, incorporating partial-ASR semantic hold further increases the F1-score to 0.4608, while maintaining a reasonable average latency of 564.0 ms. This improvement mainly comes from reducing over-segmentation caused by conversational pauses. Overall, the results show that the proposed segmentation strategy can generate usable candidate emotional segments from continuous dialogue audio and provide a practical input basis for subsequent audio-visual emotion recognition.

| Method | F1@IoU$_{0.5}$ | IoU | Latency |
|---|---|---|---|
| Fixed 800ms | 38.57 | 37.65 | 833.23ms |
| VAD-only | 45.32 | 40.69 | 544.4ms |
| Semantic-assisted | **46.08** | **41.42** | 563.96ms |

Table 1 Comparison of Automatic Segmentation Strategies

### 4.4. Comparison of hybrid fusion

This section compares the performance of unimodal models, conventional fusion strategies, and the proposed Hybrid Fusion under the five-fold LOSO setting. As shown in Table 2, the audio-only model achieves an average Accuracy of 49.19 ± 6.80% and a Macro F1-score of 49.04 ± 7.15%, whereas the visual-only model obtains 37.93 ± 4.53% Accuracy and 35.45 ± 2.06% Macro F1-score. These results indicate that, under automatically segmented inputs, the audio modality provides stronger emotion-discriminative information than the visual modality.

Among the fusion strategies, Mid Fusion achieves an average Accuracy of 52.94 ± 4.17% and a Macro F1-score of 51.81 ± 6.32%, outperforming the audio-only baseline by 3.75 and 2.77 percentage points, respectively. This suggests that, although the visual modality performs relatively weakly on its own, it can still provide complementary information when integrated at the feature level. In contrast, Late Fusion achieves 49.85 ± 4.74% Accuracy and 48.52 ± 6.52% Macro F1-score, indicating that simple decision-level fusion is less effective when the predictive strengths of the two modalities are imbalanced.

Hybrid Fusion achieves the best overall performance among the compared fusion strategies, with an average Accuracy of 53.74 ± 4.47% and a Macro F1-score of 52.61 ± 6.16%. Compared with the audio-only model, Hybrid Fusion improves Accuracy and Macro F1-score by 4.55 and 3.57 percentage points, respectively. It also outperforms Mid Fusion by 0.80 percentage points in both metrics and exceeds Late Fusion by 3.89 percentage points in Accuracy and 4.09 percentage points in Macro F1-score.

Overall, these results show that Hybrid Fusion can more effectively integrate complementary information from the audio and visual modalities than conventional Mid and Late Fusion strategies. By combining feature-level fusion with the decision information from the audio, visual, and fusion branches, Hybrid Fusion allows multiple sources of emotion evidence to contribute to the final prediction, leading to improved average recognition performance under automatically segmented audio-visual inputs.

| Method | Accuracy | F1-score |
|---|---|---|
| Unimodal 1(Audio) | 49.19 ± 6.80 | 49.04 ± 7.15 |
| Unimodal 2 (Image) | 37.93 ± 4.53 | 35.45 ± 2.06 |
| Mid | 52.94 ± 4.17 | 51.81 ± 6.32 |
| Late | 49.85 ± 4.74 | 48.52 ± 6.52 |
| Hybrid | **53.74 ± 4.47** | **52.61 ± 6.16** |

Table 2 Comparison of hybrid fusion

### 4.5. Ablation study of AGF

This section investigates the effect of incorporating the AGF module into the proposed multimodal emotion recognition framework. As shown in Table 3, the baseline Hybrid Fusion model without AGF achieves an average Accuracy of 53.74 ± 4.47% and a Macro F1-score of 52.61 ± 6.16% under the five-fold LOSO setting. After introducing AGF, the average Accuracy increases to 54.65 ± 3.98%, while the Macro F1-score improves to 53.87 ± 5.53%.

Compared with the baseline model, AGF improves Accuracy by 0.91 percentage points and Macro F1-score by 1.26 percentage points. In addition, the standard deviation decreases from 4.47% to 3.98% for Accuracy and from 6.16% to 5.53% for Macro F1-score, indicating slightly reduced performance variation across sessions.

These results suggest that audio-guided temporal fusion can improve multimodal emotion recognition under automatically segmented inputs. By using audio events to guide the search for temporally relevant visual information, AGF enables the model to better capture cross-modal temporal relationships when automatically segmented clips contain imperfect boundaries or temporal misalignment between audio and visual emotion cues. Overall, the five-fold results demonstrate that AGF provides consistent gains in average recognition performance while slightly reducing cross-session variability.

| AGF | Accuracy | F1-score |
|---|---|---|
| | 53.74 ± 4.47 | 52.61 ± 6.16 |
| √ | **54.65 ± 3.98** | **53.87 ± 5.53** |

Table 3 Ablation of AGF

### 4.6. Comparison of MNL Warmup epochs

This section investigates the effect of different warm-up epoch settings for MNL under the five-fold LOSO setting. As shown in Table 4, the warm-up duration has a clear influence on the final recognition performance. When the warm-up epoch is set to 10, the model achieves an average Accuracy of 53.72 ± 4.80% and a Macro F1-score of 53.10 ± 6.13%, which are the lowest among the evaluated settings. This suggests that introducing MNL too early may constrain learning before the modality-specific branches have developed sufficiently stable discriminative representations.

Increasing the warm-up epoch to 20 improves the average Accuracy and Macro F1-score to 55.47 ± 3.81% and 55.30 ± 5.03%, respectively. The best overall performance is obtained with a warm-up epoch of 30, reaching 57.17 ± 5.15% Accuracy and 56.79 ± 5.51% Macro F1-score. This result indicates that allowing the audio, visual, and fusion branches to establish more reliable representations before activating MNL can provide more effective negative learning guidance.

When the warm-up epoch is further increased to 40, the average Accuracy and Macro F1-score decrease to 55.22 ± 3.85% and 55.07 ± 4.12%, respectively. At 50 epochs, performance partially recovers to 56.60 ± 4.27% Accuracy and 56.38 ± 5.01% Macro F1-score, but remains slightly below the 30-epoch setting. This suggests that introducing MNL too late may reduce its opportunity to reshape the learned feature representations and decision boundaries after the model has already approached a relatively stable optimization state.

Overall, the results indicate that the effectiveness of MNL is sensitive to its intervention timing. In the present experimental setting, a 30-epoch warm-up provides the best balance between allowing the base branches to learn sufficiently discriminative representations and enabling MNL to effectively refine multimodal learning in the later training stage.

| Warmup epoch | Accuracy | F1-score |
|---|---|---|
| 10 | 53.72 ± 4.80 | 53.10 ± 6.13 |
| 20 | 55.47 ± 3.81 | 55.30 ± 5.03 |
| 30 | **57.17 ± 5.15** | **56.79 ± 5.51** |
| 40 | 55.22 ± 3.85 | 55.07 ± 4.12 |
| 50 | 56.60 ± 4.27 | 56.38 ± 5.01 |

Table 4 MNL warmup epochs

### 4.7. Ablation of PDF

This section investigates the effect of the PDF module on recognition performance and noise robustness under the five-fold LOSO setting. As shown in Table 5, the baseline model without PDF achieves an average Macro F1-score of 56.79 ± 5.51% under the clean condition. After introducing PDF, the Macro F1-score increases to 57.40 ± 5.87%, corresponding to an improvement of 0.61 percentage points. This result suggests that PDF can provide additional performance gains even under clean conditions by dynamically adjusting the contributions of different prediction branches according to their estimated reliability.

When noise is introduced, the benefit of PDF becomes more pronounced. Under the moderate noise condition ($\lambda = 5$), the average Macro F1-score improves from 53.07 ± 4.44% to 55.21 ± 5.23%, yielding a gain of 2.13 percentage points. Under the higher noise condition ($\lambda = 10$), the Macro F1-score increases from 50.60 ± 3.73% to 52.58 ± 4.52%, corresponding to an improvement of 1.99 percentage points. These gains are larger than those observed under the clean condition, indicating that PDF is particularly beneficial when modality quality is degraded.

Overall, the results demonstrate that PDF can mitigate the performance degradation caused by noisy or unreliable modality inputs. By estimating branch reliability and dynamically adjusting decision-level contributions, PDF reduces the influence of less reliable predictions and helps preserve more robust multimodal emotion recognition under varying modality quality.

| Method | $\lambda = 0$ | $\lambda = 5$ | $\lambda = 10$ |
|---|---|---|---|
| | 56.79 ± 5.51 | 53.07 ± 4.44 | 50.60 ± 3.73 |
| +PDF<br>Δ | **57.40 ± 5.87**<br>+0.61 | **55.21 ± 5.23**<br>+2.13 | **52.58 ± 4.52**<br>+1.99 |

Table 5 Noise Robustness Analysis of PDF

### 4.8. Comparison of other Models

This section compares the proposed AGF-Net with existing multimodal emotion recognition methods under clean and noisy conditions. As shown in Table 6, under the clean setting ($\lambda = 0$), AGF-Net achieves an Accuracy of **64.92%** and an F1-score of **65.47%**. Although its performance is slightly lower than that of Social-MAE, which achieves 66.72% Accuracy and 67.89% F1-score, AGF-Net outperforms AVF-MAE and VAVL, demonstrating competitive emotion recognition performance under clean conditions.

When noise is introduced, AGF-Net maintains strong recognition performance. Under the moderate noise condition ($\lambda = 5$), AGF-Net achieves **62.13% Accuracy** and **63.08% F1-score**. Its Accuracy is comparable to that of Social-MAE (62.17%), while its F1-score is the highest among all compared methods, exceeding Social-MAE, AVF-MAE, and VAVL by **0.52, 4.15, and 7.89 percentage points**, respectively. This result indicates that AGF-Net is able to maintain effective multimodal emotion recognition when the quality of one modality is degraded.

Under the more challenging noise condition ($\lambda = 10$), AGF-Net achieves **58.19% Accuracy** and **59.41% F1-score**. Although its Accuracy is lower than those of AVF-MAE and Social-MAE, its F1-score remains highly competitive and is only 0.20 percentage points lower than that of Social-MAE (59.61%), while still outperforming AVF-MAE and VAVL. These results suggest that the proposed framework retains effective classification capability even under substantial modality degradation.

Overall, AGF-Net does not achieve the highest score for every metric under all conditions; nevertheless, it consistently provides competitive performance across different noise levels. In particular, it achieves the highest F1-score under the moderate noise condition and remains close to the best-performing method under severe noise. These results demonstrate that the proposed AGF, MNL, and PDF mechanisms enable AGF-Net to maintain robust multimodal emotion recognition performance under varying modality quality and noise conditions.

| Method | $\lambda = 0$ Acc/F1 | $\lambda = 5$ Acc/F1 | $\lambda = 10$ Acc/F1 |
|---|---|---|---|
| AGF-Net(Ours) | 64.92/65.47 | 62.13/**63.08** | **58.19/59.41** |
| Social-MAE[34] | **66.72/67.89** | **62.17**/62.56 | 58.68/59.61 |
| AVF-MAE[35] | 59.51/59.78 | 59.02/58.93 | 60.00/59.25 |
| VAVL[36] | 55.41/57.17 | 53.44/55.19 | 50.16/50.85 |

Table 6 Comparison of other models

### 4.9. Experiment Summary

Section 4.3 evaluates the automatic utterance segmentation strategy, showing that the proposed segmentation process can generate usable candidate clips for subsequent audio-visual emotion recognition. Section 4.4 compares Hybrid Fusion with unimodal models and conventional fusion strategies under the five-fold LOSO setting. Hybrid Fusion achieves the best overall performance among the evaluated fusion methods, with an average Accuracy of 53.74 ± 4.47% and a Macro F1-score of 52.61 ± 6.16%. In Section 4.5, the ablation study shows that incorporating AGF further improves the average Accuracy to 54.65 ± 3.98% and the Macro F1-score to 53.87 ± 5.53%, corresponding to gains of 0.91 and 1.26 percentage points, respectively.

Section 4.6 examines the warm-up setting for MNL. Among the evaluated settings, a 30-epoch warm-up achieves the best average test performance, reaching 57.17 ± 5.15% Accuracy and 56.79 ± 5.51% Macro F1-score. Section 4.7 further evaluates PDF under clean and noisy conditions. Under the clean condition, PDF improves the Macro F1-score from 56.79 ± 5.51% to 57.40 ± 5.87%. Under noise levels of ($\lambda = 5$) and ($\lambda = 10$), the Macro F1-score improves from 53.07 ± 4.44% to 55.21 ± 5.23% and from 50.60 ± 3.73% to 52.58 ± 4.52%, respectively, indicating that PDF can mitigate performance degradation caused by degraded modality quality.

Finally, Section 4.8 compares the proposed AGF-Net with existing multimodal emotion recognition models under clean and noisy conditions. AGF-Net demonstrates competitive performance across different noise levels, achieving 64.92% Accuracy and 65.47% F1-score under the clean setting ($\lambda = 0$), 62.13% Accuracy and 63.08% F1-score under the moderate noise setting ($\lambda = 5$), and 58.19% Accuracy and 59.41% F1-score under the higher noise setting ($\lambda = 10$). In particular, AGF-Net achieves the highest F1-score under the moderate noise condition and remains close to the best-performing method under the higher noise condition. Overall, these results demonstrate that the proposed framework can effectively perform audio-visual emotion recognition on automatically segmented inputs while maintaining competitive and robust performance under varying modality quality.

## 5. Result

The proposed AGF-Net achieves 64.92% Accuracy and 65.47% F1-score under the clean setting. Under noisy conditions, it maintains 62.13% Accuracy and 63.08% F1-score at ($\lambda = 5$), and 58.19% Accuracy and 59.41% F1-score at($\lambda = 10$). In the ablation studies under the five-fold LOSO setting, incorporating the AGF module improves the average Accuracy from 53.74 ± 4.47% to 54.65 ± 3.98% and the Macro F1-score from 52.61 ± 6.16% to 53.87 ± 5.53%. For MNL, the 30-epoch warm-up achieves the best average test performance, reaching 57.17 ± 5.15% Accuracy and 56.79 ± 5.51% Macro F1-score. Finally, introducing PDF further improves the clean-condition Macro F1-score from 56.79 ± 5.51% to 57.40 ± 5.87%, while providing larger gains under noisy conditions, reaching 55.21 ± 5.23% at($\lambda = 5$)and 52.58 ± 4.52% at($\lambda = 10$).

## 6. Future Work

This work can be further extended in three directions. First, additional auxiliary modalities, such as textual, semantic, or physiological signals, can be incorporated to complement audio and visual information and further improve the recognition of complex emotional states. Second, contextual information can be further utilized to assist emotion prediction, such as neighboring utterances, dialogue history, or short-term memory mechanisms, allowing the model to consider emotional changes across continuous conversations and reduce the impact of insufficient information or incomplete automatic segmentation in a single segment. Third, the proposed method can be evaluated on other audio-visual emotion recognition datasets to further verify its generalization ability across different data distributions, speakers, languages, and interaction scenarios.